\documentclass{emulateapj}
\shortauthors{Wang \& Chakrabarty}

\newcommand{\shump}{4U~1820$-$30}

\begin{document}

\title{Discovery of A 693.5 s Period in the X-Ray Binary 4U~1820$-$30: A Superhump Interpretation} 

\author{Zhongxiang Wang \& Deepto Chakarbarty}
\affil{Shanghai Astronomical Observatory, Chinese Academy of Sciences,\\
80 Nandan Road, Shanghai 200030, China}
\affil{Department of Physics and 
Kavli Institute for Astrophysics and Space Research,\\
Massachusetts Institute of Technology, Cambridge, MA 02139, USA}


\begin{abstract}
The X-ray source \shump\ in the globular cluster NGC~6624
is known as the most compact binary among the identified X-ray binaries.
Having an orbital period of 685.0~s, the source consists of
a neutron star primary and likely a 0.06--0.08 $M_{\sun}$
white dwarf secondary. Here we report on far-ultraviolet (FUV) 
observations of this X-ray binary, made with the Space Telescope 
Imaging Spectrograph on board the {\em Hubble Space Telescope}. 
From our Fourier spectral analysis of the FUV timing data, we obtain 
a period of $693.5\pm 1.3$~s, which is significantly different 
from the orbital period. The light curve folded at this period 
can be described by a sinusoid, with a fractional semiamplitude of 6.3\% 
and the phase zero (maximum of the sinusoid) at 
MJD 50886.015384$\pm$0.000043 (TDB).
While the discovered FUV period may be consistent with a hierarchical 
triple system model that was previously considered for 4U 1820$-$30, 
we suggest that it could instead be the indication of 
superhump modulation, which arises from an eccentric accretion 
disk in the binary. The X-ray and FUV periods would be
the orbital and superhump periods, respectively, indicating a 1\% superhump
excess and a white-dwarf/neutron-star mass ratio around 0.06.
Considering 4U 1820$-$30 as a superhump source, we discuss the implications. 

\end{abstract}


\section{INTRODUCTION}

Among the known X-ray binaries (XRBs) that consist
of an accreting neutron star (NS) or black hole primary,
the NS binary 4U~1820$-$30 in the globular
cluster NGC~6624 is known as the most compact binary (e.g., \citealt{rk09}): 
its orbital period $P_{\rm orb}= 685.0119 \pm 0.0001$~s 
\citep*{spw87,cg01}. 
Considering this short period, the Roche-lobe filling
companion in the binary can be estimated to be a 0.06--0.08 $M_{\sun}$, 
hydrogen-exhausted white dwarf (WD; \citealt{rap+87}).
The X-ray source shows a long-term modulation 
with a period of approximately 171 days \citep{pt84b, cg01, zwg07},
which was interpreted as an indication that this source is a hierarchical 
triple system \citep{gri88, cg01}.
In this triple system model,  the NS-WD binary is the inner 
binary with orbital period $P_{\rm inner}\approx 685$ s,  
and a second companion, orbiting the inner system
with a period of $P_{\rm outer} \sim 1$ day, induces an inner-binary 
eccentricity variation with the 171 day period.  Consequently, 
the mass accretion rate and X-ray luminosity of the X-ray source 
modulate with the same period.

In this paper, we report on the detection of a periodicity in \shump\  
from far-ultraviolet (FUV) observations, made with the 
{\em Hubble Space Telescope} ({\em HST}).
This FUV period is about 1\% longer than the X-ray period, 
a difference that approximately fits 
the triple system model: one of the two periods, either the X-ray or FUV, 
is $P_{\rm inner}$,  while the other one is the beat period between 
$P_{\rm inner}$ and $P_{\rm outer}$.  However,  since 
a very similar source, the XRB 4U~1916$-$05 (which has X-ray
and optical periods of 50.00 min and 50.46 min, respectively),
has been verified as a superhump source by the detection of negative 
superhumps in X-ray observations (\citealt{ret+02} and reference
therein), the slightly longer FUV period detected in \shump\  
could well be the indication of the binary also being a superhump source.
Additionally, a long term modulation (199~day)
in the X-ray light curve of 4U~1916$-$05 had also been reported
\citep{pt84a, sl92}, although \citet{hom+01} only found a likely 83 day 
modulation. 
Due to the similarities between the two sources, we thus 
suggest a superhump model for 4U~1820$-$30 in this paper. 
\begin{figure*}
\begin{center}
\includegraphics[scale=0.82]{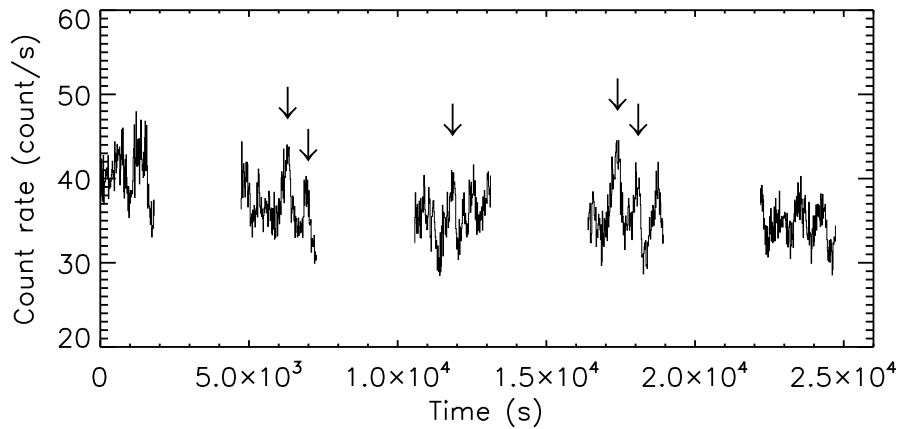}
\figcaption{{\em HST}/STIS FUV light curve of 4U 1820$-$30. For clarity, 
the photon counts are binned in 16 s time intervals. A  
modulation is clearly visible in the light curve. 
The modulation peaks that are used
in the data analysis in \S~\ref{sec:res} are marked by arrows.
\label{fig:lc}} 
\end{center}
\end{figure*}

Superhumps are periodic modulations observed in interacting 
binary systems with periods a few percent longer than their
orbital periods. This phenomenon was first discovered in
the super-outbursts of the dwarf nova VW Hyi \citep{vog74, war75},
and since then, it has
been commonly detected in short-period cataclysmic variables 
(e.g., \citealt{pat98, ski+99}).
It is generally believed that superhumps arise from an
elliptical accretion disk, which is developed when the disk extends
beyond the 3:1 resonance radius and precesses in the inertial frame 
due to the tidal force of a secondary star (e.g., \citealt{wk91}).
The resonance condition---the mass ratio of
a secondary to a primary $q\lesssim 0.33$---appears
to also work in X-ray binaries, 
as superhumps have been detected in both black hole soft X-ray 
transients (SXTs; \citealt{oc96})
and NS XRBs 4U~1916$-$053 and GX~9+9 \citep{has+01}.
Indeed, Haswell et al. (2001) have further suggested that NS low-mass
XRBs with orbital periods below $\sim 4.2$ hr are potential 
superhump sources.

Previous ultraviolet (UV) observations of \shump,
made with the former {\em HST} instrument 
Faint Object Spectrograph (FOS) in {\tt RAPID} mode 
(in a wavelength range of 1265--2510 \AA),
revealed a periodic modulation with a 16\% amplitude \citep{and+97}.
Although the reported UV period, 687.6$\pm$2.4~s, is consistent with
the X-ray period, the accuracy of the timing result should have been hampered
by several uncertainties on FOS timing in {\tt RAPID} mode (for details, see 
FOS Instrument Science Report CAL/FOS \#124, 
\#150)\footnote{\url{http:://www.stecf.org/poa/FOS/fos\_bib.html}}.

\section{OBSERVATIONS AND DATA REDUCTION}    

The {\em HST} observations were made on 1998 March 14 using the Space 
Telescope Imaging Spectrograph (STIS; \citealt{woo+98}).
A low-resolution grating G140L was used with the solar-blind CsI 
multi-anode microchannel array (STIS/FUV-MAMA) detector, providing 
an FUV wavelength coverage of 1150-1730 \AA\  and a plate scale
of 0.025\arcsec/pixel along the spatial direction.  
The aperture used was a 52\arcsec$\times$0.5\arcsec\ long slit. 
The data were taken in {\tt TIMETAG} mode,  recording 
the detector position and arriving time of each detected photon.
The timing resolution was 0.125 ms.
We obtained the data from the {\em HST} archive service. 
The datasets consist of five exposures with a total exposure time of  
12.1~ks:  the first dataset had an exposure time of 1830~s,  
while the exposure time for each of the other four was 2580~s. 
The total time span was $T_{\rm s}\approx 24.7$ ks ($\simeq 7$ hrs). 

Using the IRAF task {\tt odelaytime} in the {\em HST} calibration 
package {\tt STIS}, we applied a barycenter correction to the measured
arrival times.
The {\em HST} ephemeris (ORX) file at the observation time
was obtained using the {\tt Starview} program.  A box region with 
a 30-pixel (=0\farcs75) width in the spatial direction, which well covered
the target's spatial profile, was used for collecting the target's counts.
We excluded the counts within the Ly$\alpha$ 
$\lambda$1216 \AA\ line, for which the source was strong background emission
\citep{and+97}.
The resulting light curve of \shump\ is shown in Figure~1. 
A modulation is clearly visible in the light curve.
The contamination
from the background was limited: in a nearby region with
the same size to that of the target region, we found a background
count rate of only 1.3 counts s$^{-1}$ (compared to 30--45 counts s$^{-1}$ 
in the target region).  
\begin{center}
\includegraphics[scale=0.7]{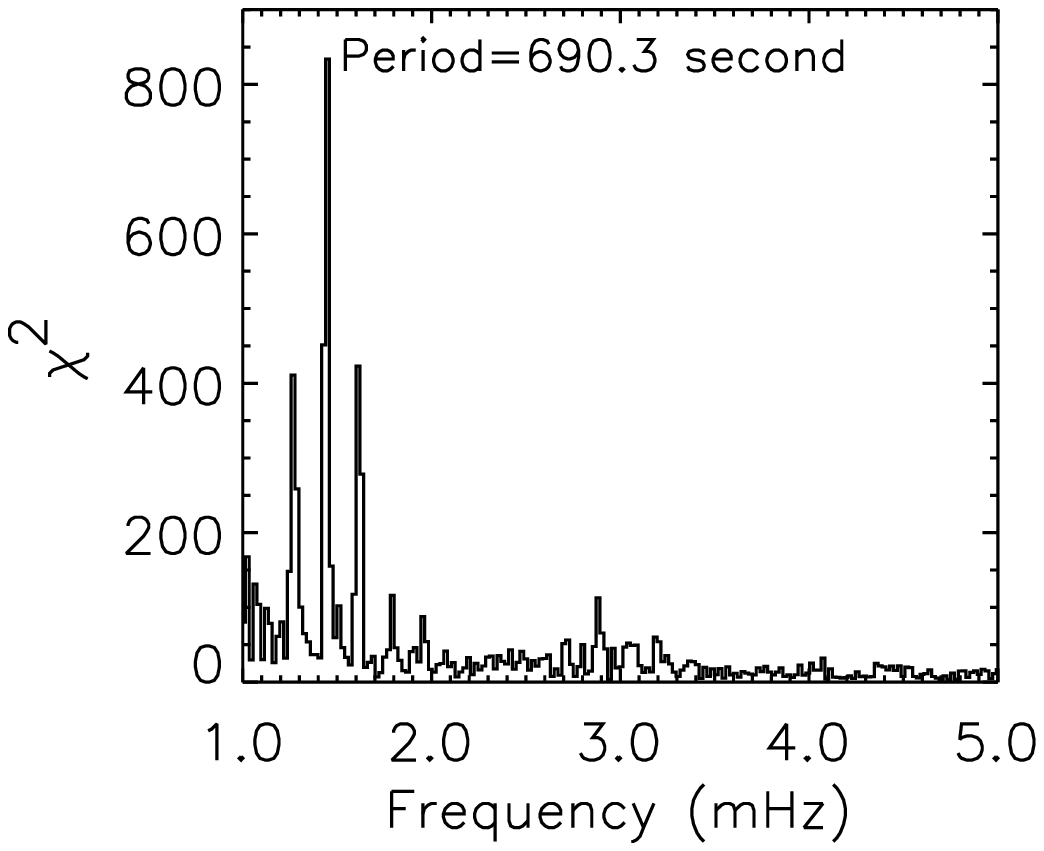}
\figcaption{Result of the epoch-folding analysis. A period of
690.3 s is clearly detected. Two major sidebands are artifacts 
due to the observation gaps.
\label{fig:efold}}
\end{center}

\section{RESULTS}
\label{sec:res}

We first used an epoch-folding technique
(Leahy et al.~1983) to search for periodicity, with the data folded to 
10 phase intervals per period.
The resulting $\chi^2$ values are shown in Figure~2, clearly
indicating the detection of a periodicity. The period is 
$P = 690.3$ s, 5.3~s longer than the 685.0~s X-ray
period. However, based on the epoch-folding technique, 
the frequency spacing was 
$1/2T_{\rm s} = 2.02\times 10^{-5}$ Hz, corresponding to 
a temporal spacing of 9.6~s near 690~s. As a result, the spacing is not
sufficient to determine the period difference
between 690.3~s and 685.0~s. 

To more accurately determine the period, we 10 times
over-sampled our data and used a discrete Fourier transform technique to
construct an over-resolved power spectrum (e.g., \citealt{cha98, rem02}).
The original data were binned evenly in 1-s time intervals for the Fourier 
transform.
In Figure~3, the over-resolved power spectrum in the vicinity of
the main power peak is shown. We fit the peak with a Gaussian 
to obtain the peak frequency, and found $f = 1.4442\pm 0.0027$ mHz, 
corresponding to a period of $693.5\pm 1.3$~s, where the standard 1$\sigma$
uncertainties for signal frequency in an over-resolved power spectrum 
are given  \citep{mn76,rem02}.
This indicates that the period difference between the FUV and X-ray periods 
is significant (approximately 6$\sigma_{P}$ confidence). 
\begin{center}
\includegraphics[scale=0.55]{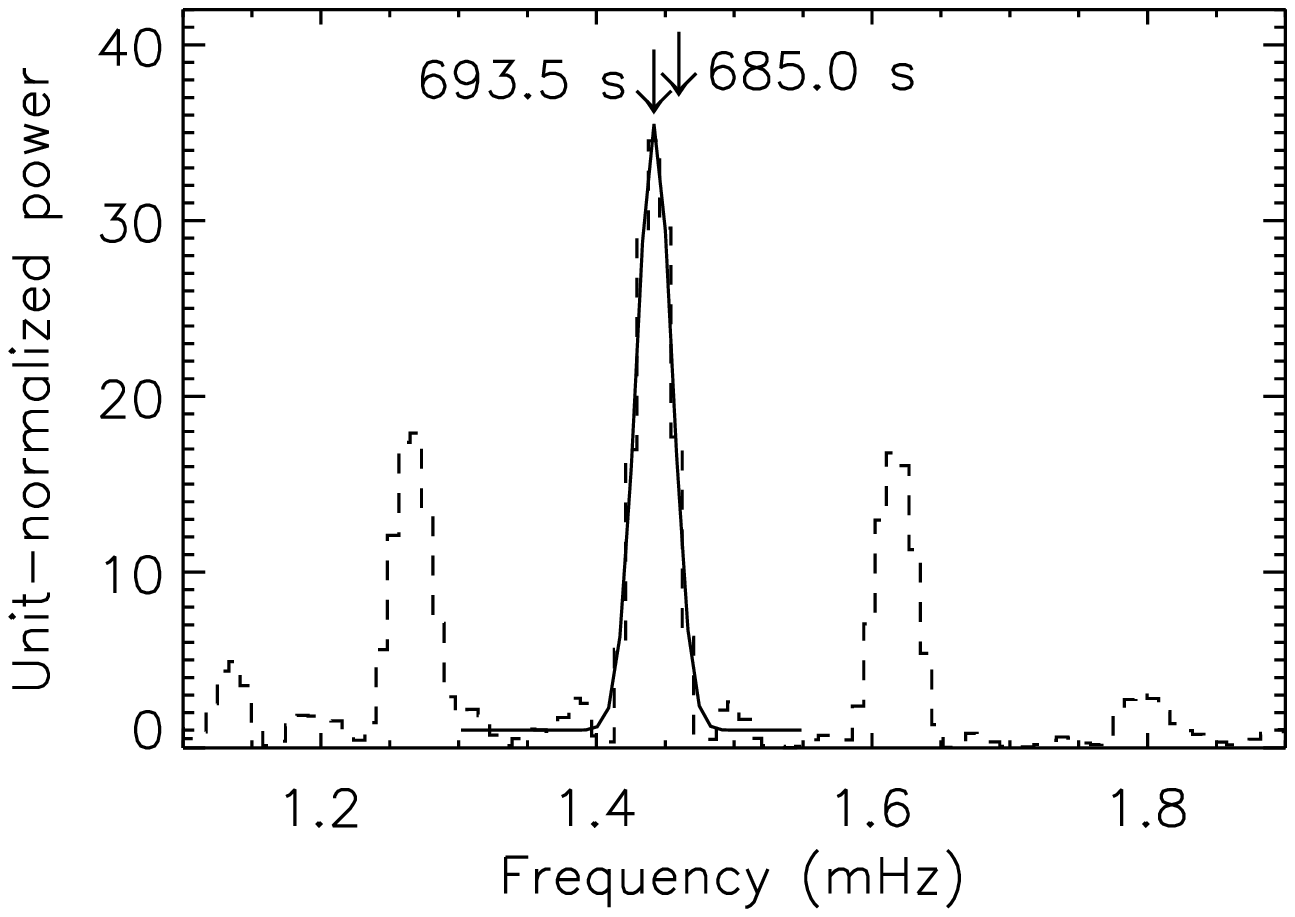}
\figcaption{A 10 times over-resolved power spectrum (dashed curve)
in the vicinity of
the main power peak,  resulting from the discrete Fourier transform. 
Two major sidebands
are artifacts due to the observation gaps. 
Fitting the peak with a Gaussian function (solid curve), the peak frequency
is found to be $f = 1.4442\pm0.0027$ mHz (or 693.5$\pm1.3$~s). 
The frequency position of the X-ray period 685.0~s
is also indicated.
\label{fig:npx}}
\end{center}

Careful examination of the light curves in Figure~\ref{fig:lc} reveals that  
the strongest modulation peaks are asymmetric and non-sinusoidal in shape. 
As an additional check on our determination of the modulation period, 
we also used a phase-dispersion minimization technique \citep{ste78}. 
This technique works well for cases of non-sinusoidal variation contained
in a few irregularly spaced observations.
The resulting periodogram in the vicinity of the minimum $\Theta$ 
statistic is shown in Figure~\ref{fig:pdm}. 
Fitting the region near the minimum
with a parabola \citep{ste78}, we found $P=693.3$ s. The period value,
again deviating from 685.0~s, is consistent with
that obtained from the Fourier transform.

In order to quantify the overall periodic modulation in the light curve and 
thereby compare it with that obtained by \citet{and+97}, 
we binned the data in 10 s time intervals and obtained
1215 data points. The data points in each exposure interval
were subtracted and normalized by the average,
and then folded at the period of 693.5~s. The folded data points, 
shown in the top panel of Figure~\ref{fig:fold}, are scattered in 
a relatively large range. 
For example, further binning the folded data points into 10 phase bins, 
the standard deviations of the bins were found to be between 5.9--8.4\%.
Large scattering in optical light curves of XRBs is commonly seen 
(e.g., \citealt{wang+09}).  For our case, 
we note that the modulation amplitude has significant variations in 
the data (Figure~\ref{fig:lc}). 
For example, in several regions, the modulation is barely visible.
These amplitude variations smear the periodic modulation and cause
large scattering in the folded light curve.
\begin{center}
\includegraphics[scale=0.75]{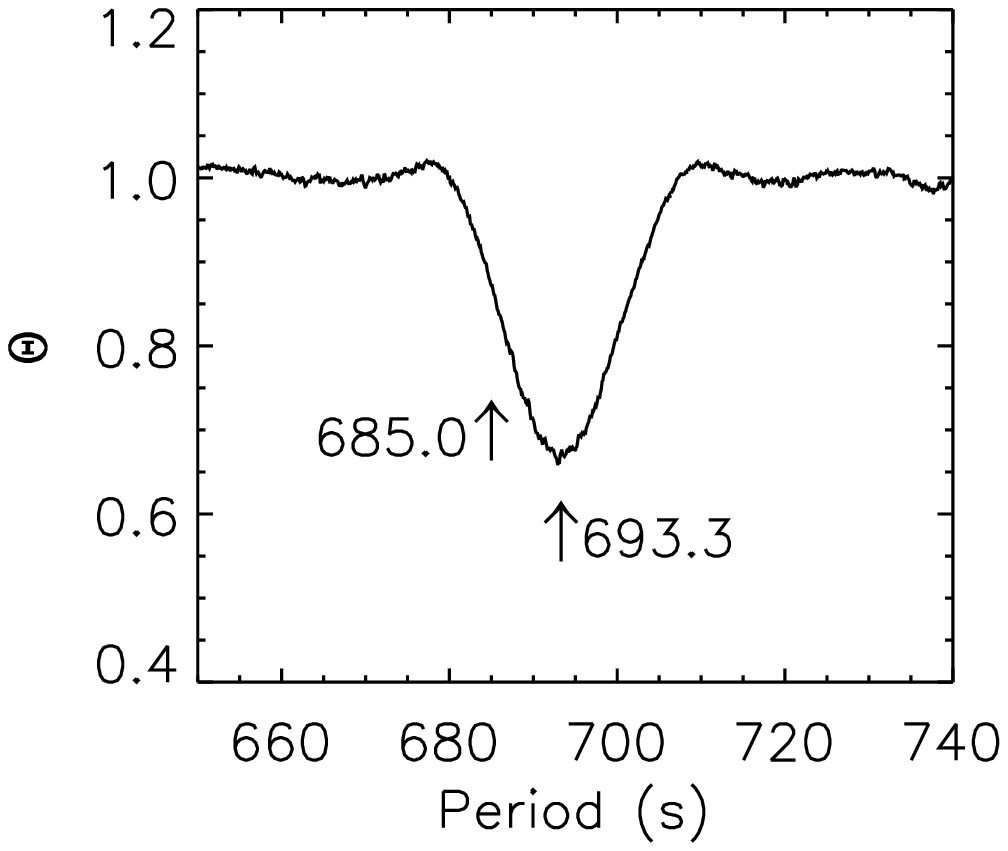}
\figcaption{Phase dispersion minimization periodogram in the vicinity 
of the minimum $\Theta$ statistic.
The positions of the minimum $\Theta$ statistic and
X-ray period 685.0 s are marked by arrows to indicate their difference.
\label{fig:pdm}}
\end{center}

Following \citet{and+97}, we fitted a sinusoid to the 
1215 data points, with the period fixed at 693.5 s. 
The best-fit has reduced $\chi^2= 1.9$ for 1212 degrees of freedom,
reflecting large scattering of the data points.
From the best-fit, we found a semiamplitude of 6.3\%.  
The time at the maximum of the sinusoidal fit (phase zero) was
MJD 50886.015384$\pm$0.000043 (TDB) at the solar system
barycenter.
This semiamplitude we obtained is 2\% lower
than that in the UV range reported by \citet{and+97}. 
This difference could be due to the different wavelength coverages,
because while the time spans of the UV and FUV data were approximately equal,
the former covered a wavelength range twice as large as the latter. 
In addition, the UV observations were made in 1996, two years earlier
than the FUV observations. Therefore the difference
could also be due to intrinsic variations in UV emission from the source.

The strong modulation peaks are asymmetric.  Here we
folded 5 such peaks to show the asymmetry feature. 
In Figure~\ref{fig:lc}, the peaks that were used
in folding are marked by arrows. The folded light curve is shown 
in the bottom panel of Figure~5. 
As can be seen, the light curve has a slow rise and a fast decline,
with a semiamplitude of $\simeq$10\%. This asymmetric modulation is 
different from those sinusoidal or double-hump modulations that
arise from companion stars and normally seen in XRBs (e.g., \citealt{vm95}).
Rather, it resembles those modulations seen in superhump 
sources (e.g., \citealt{rlo97}).

\section{DISCUSSION} 

Using different techniques in our timing analysis, we have discovered 
a period that is significantly different from the X-ray period.  
X-ray observations of \shump\  by various X-ray observatories have consistently 
obtained the 685.0~s period over the past two decades (e.g., \citealt*{smm87};
\citealt*{mrg88}; \citealt{san+89}; \citealt{tan+91}; \citealt{vdk+93b};
\citealt{cg01}), demonstrating the stableness of 
the X-ray period. Also even though an X-ray period derivative $\dot{P}$ 
has been detected \citep{tan+91, vdk+93b, cg01}, its negative value, 
$\dot{P}/P \approx$ $-$(2--5)$\times 10^{-8}$ yr$^{-1}$,  not only implies
a decrease of the period as a function of time,  but also is too small
($\Delta P \leq $0.035 ms yr$^{-1}$) 
to account for any period changes comparable to
the 8.5~s difference between the X-ray and FUV periods. 
As STIS may be repaired in the future, our result could be
confirmed by further observations. We would use a phase-coherent
timing technique to build a long time span from several short observations,
thus allowing a very accurate determination of the FUV period 
(e.g., \citealt{wang+09}).
\begin{center}
\includegraphics[scale=0.7]{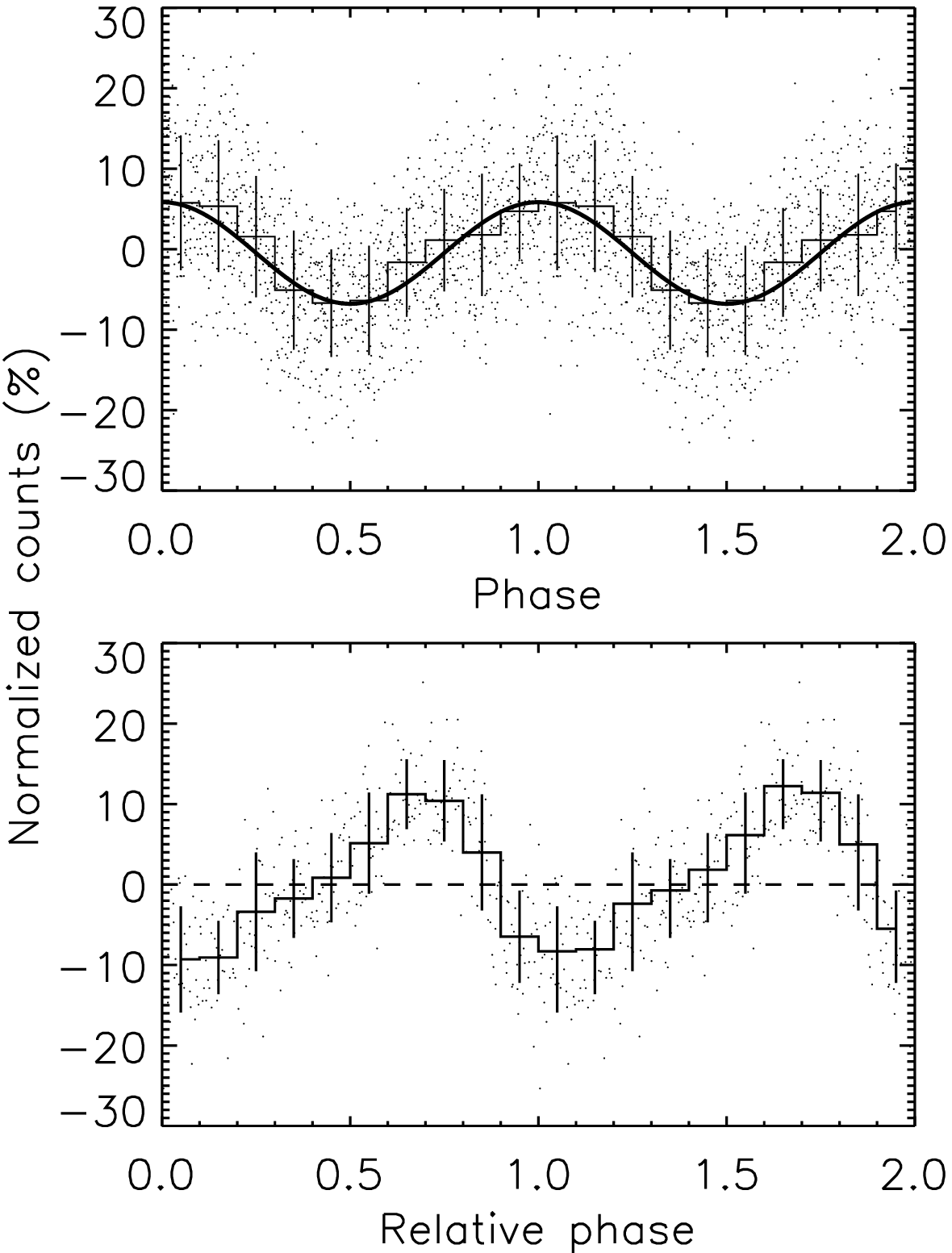}
\figcaption{
{\it Top panel:} normalized photon counts (relative to the mean) folded
with the 693.5-s period. Two cycles are displayed for clarity.  The 
solid curve indicates the best-fit sinusoid that has a semiamplitude
of 6.3\%. However, the data points are scattered in a relatively large
range, indicated by the 5.9--8.4\% standard deviations (error bars) of the bins
(histogram line); each bin is the average of the data points within
0.1 phase. {\it Bottom panel:} same as the top panel, but only 
photon counts contained in the strong modulation peaks 
(see Figure~\ref{fig:lc}) are used in folding.
\label{fig:fold}}
\end{center}

The discovered period could be additional evidence for 
the triple system model, which has been suggested
for \shump\  (\citealt{gri88, cg01}).
Since the source is located near the center of the
globular cluster \citep{kin+93}, the existence of triple systems
is expected due to binary-binary interactions (e.g., \citealt*{rmh95}).
Indeed, one triple system, known as PSR B1620$-$26,
has been found in the globular cluster M4, although
this source may not represent a prototypic triple system
in a globular cluster \citep{st05}. The second companion
around the system's inner NS-WD binary is a Jovian planet \citep{sig+03}.

However, given the similarities between the XRBs 4U 1820$-$30 and 4U 1916$-$053,
it is likely that 4U 1820$-$30 is a superhump source with
the X-ray and FUV periods being $P_{\rm orb}$
and the superhump period $P_{\rm sh}$, respectively.
Since the orbital periodicity is indicated by dips in the X-ray light curve
of the source, this XRB probably has an inclination angle of $\sim$60\arcdeg\ 
\citep{vm95}.
With this angle, the UV modulation that is caused by X-ray irradiation
of the companion star has been estimated to be
at the 5\% level of the persistent emission from the 
accretion disk \citep{ak93}.
The superhump modulation, presumably reflecting the area changes of
the X-ray irradiated accretion disk, could be comparable
but slightly stronger \citep{has+01}. Therefore the FUV modulation we detected
could arise from both the companion and disk, but with the latter 
slightly dominant over the former. This might explain large scattering
seen in the folded light curve. In addition, the strong modulation peaks in 
the light curve
appears to be asymmetric, resembling those seen in superhump sources.
This asymmetry feature in the modulation provides additional support to 
our suggestion that \shump\ is a superhump source.
The origin of the long-term 171 day variation is not clear.
Simply by comparison, it would be the so-called ``superorbital" modulation 
since long-term periodicities have been seen in a few other XRBs.
This type of modulations has been suggested 
to be caused by the nodal precession of a warped disk (e.g., \citealt{cla+03}).
However, critical questions remain to be answered for the warping disk scenario 
(e.g., \citealt{cla03,zdz07}).  

In the superhump model, the superhump excess is defined as
$\epsilon = (P_{\rm sh} - P_{\rm orb})/P_{\rm orb}$.
For \shump, we obtain $\epsilon\approx 0.012\pm0.002$,
similar to those values found for 4U~1916$-$053 (\citealt{has+01}) 
and SXTs \citep{oc96}.
Based on a relation between the superhump excess
and mass ratio $q$ \citep{pat+05},
$\epsilon =0.18q+0.29q^2$,
it can be found that $q\approx 0.06$ for \shump, implying
a companion mass of about $0.08\ M_{\sun}$
(a $1.4\ M_{\sun}$ standard neutron star mass is assumed).
This mass value is consistent with the mass range 
estimated for \shump\  in a standard scenario (i.e., the mass transfer
in the binary system is driven by gravitational radiation; 
\citealt{rap+87}). 

It is interesting to note that the $\epsilon$--$q$ relation 
could be used to set a constraint on the mass of a neutron star if 
the mass of a companion could be estimated from other measurements. 
Currently, three distinct binary evolution channels are thought to form 
ultracompact binaries (whose orbital periods are less than $\simeq$80 min;
for ultracompact binary formation and evolution, see \citealt{del+07} and 
references therein). Assuming the WD channel for the formation of \shump,
the helium WD in this XRB would likely have a mass 
of 0.06--0.1 $M_{\sun}$ at the present time, having evolved from
an initial mass of 0.1--0.3 $M_{\sun}$ at the beginning of the XRB phase
(\citealt{del+07}; Deloye 2009, private communication). 
The $q$ value estimated from the superhump period suggests that the 
accreting neutron star in \shump\ has a mass value not much larger than  
$1.4\  M_{\sun}$, the canonical neutron star mass.

The observed negative $\dot{P}$ has been a puzzling feature
in \shump\  \citep{vdk+93b, cg01},
since in the standard scenario, a positive $\dot{P}$ for 4U 1820$-$30 is 
expected \citep{rap+87}.
The negative value could be caused by the gravitational acceleration
of the globular cluster \citep{tan+91, vdk+93b, kin+93}. 
Alternatively as suggested by \citet{vdk+93a}, 
the period changes could be caused by azimuthal variations of the
impact point in the accretion disk (at which the gas flow from
the companion star impinges on the disk), and the azimuthal variations could
be induced by the precession of an elliptical accretion disk. 
As a result, the phase of the observed X-ray light curve would be shifted and 
thus there might not be any changes for the true orbital period (however,
see \citealt{cg01} for arguments against this scenario).
Our identification of \shump\ as a superhump source may provide
a support to this scenario, since the accretion disk in a superhump source
is expected to be elliptical.

In order to confirm whether 4U 1820$-$30 is a superhump system,  
critical evidence such as the detection of
negative superhumps is required \citep{ret+02}.  
According to the empirical relation among the periods of 
the negative and positive superhump and orbit 
given by \citet{ors09}, the negative
superhump period should be $\sim$678 s.
We may also search for a beat period between 
the X-ray and FUV/optical periods from 4U~1820$-$30, since such 
a period was found from the X-ray light curve of 4U~1916$-$05 (3.90 days; \citealt*{cgb01}; \citealt{hom+01}).
This beat period 
indicates that an accretion disk precesses with a period
$P_{\rm prec} = (1/P_{\rm orb} - 1/P_{\rm sh})^{-1}$.
For 4U~1820$-$30, we estimate $P_{\rm prec}\approx 0.7\pm 0.1$ day.
However,  because the X-ray timing observations of \shump\ were mainly
made with {\em Rossi X-Ray Timing Explorer} ({\em RXTE}), 
searches for such a period are difficult because of
$\sim$1 day observation gaps of {\em RXTE} data.
We note that our FUV observations were made near the bottom flux 
of the 171 day periodic modulation.
In the superhump scenario, 
both the superhump and superorbital modulations are supposed to arise 
from the accretion disk in 4U 1820$-$30. 
A detection of any correlations between these two
modulations, such as variations of the superhump period and light curve as
a function of superorbital phase \citep{vdk+93a}, 
would verify the disk origin of the modulation and thus support
that 4U 1820$-$30 is a superhump source.

\acknowledgements

We thank Jennifer Sokoloski for advice on timing analysis,
Michael Nowak for a helpful discussion, and the anonymous referee for valuable
suggestions. We also thank Lars Bildsten, Chris Deloye, and Andrzej Zdziarski 
for useful comments on an earlier version of the manuscript.

\bibliographystyle{apj}

\end{document}